\documentclass{article}

\usepackage{arxiv}

\usepackage[utf8]{inputenc} 
\usepackage[T1]{fontenc}    
\usepackage{hyperref}       
\usepackage{url}            
\usepackage{booktabs}       
\usepackage{amsfonts}       
\usepackage{nicefrac}       
\usepackage{microtype}      
\usepackage{lipsum}
\usepackage{subcaption}
\usepackage{amsmath,amsfonts}
\usepackage{xcolor}
\usepackage{graphicx}
\usepackage{comment}
\usepackage{algorithm}
\usepackage{algpseudocode}
\graphicspath{ {./images/} }

\title{Spectral Feature Extraction for Robust Network Intrusion Detection Using MFCCs}

\author{
 HyeYoung Lee \\
  SPILab Corporation\\
  Ulsan \\
  \texttt{uohesha@korea.ac.kr} \\
   \And
 Muhammad Nadeem \\
  SPILab Corporation\\
  Ulsan \\
  \texttt{mnadeem@spilab.kr}\\
  \And
 Pavel Tsoi \\
 SPILab Corporation \\
 Ulsan \\
 \texttt{pauts@spilab.kr} \\
}

\begin{document}
\maketitle
\begin{abstract}
The rapid expansion of Internet of Things (IoT) networks has led to a surge in security vulnerabilities, emphasizing the critical need for robust anomaly detection and classification techniques. In this work, we propose a novel approach for identifying anomalies in IoT network traffic by leveraging the Mel-frequency cepstral coefficients (MFCC) and ResNet-18, a deep learning model known for its effectiveness in feature extraction and image-based tasks. Learnable MFCCs enable adaptive spectral feature representation, capturing the temporal patterns inherent in network traffic more effectively than traditional fixed MFCCs. We demonstrate that transforming raw signals into MFCCs maps the data into a higher-dimensional space, enhancing class separability and enabling more effective multiclass classification. Our approach combines the strengths of MFCCs with the robust feature extraction capabilities of ResNet-18, offering a powerful framework for anomaly detection. The proposed model is evaluated on three widely used IoT intrusion detection datasets: CICIoT2023, NSL-KDD, and IoTID20. The experimental results highlight the potential of integrating adaptive signal processing techniques with deep learning architectures to achieve robust and scalable anomaly detection in heterogeneous IoT network landscapes.
\end{abstract}

\section{Introduction}
\label{introduction}
In recent years, the proliferation of sensors and advancements in wireless communication technologies have fueled the rapid expansion of Internet of Things (IoT) networks. These networks have become a cornerstone for a wide range of modern applications, including smart homes, healthcare, smart cities, smart grids, intelligent transportation systems, and numerous other innovations \cite{ma2020towards}. IoT networks consist of interconnected devices equipped with various sensors, actuators, storage components, and computing and communication functionalities. These devices are responsible for collecting data and transmitting them over the traditional Internet \cite{mehmood2017internet}. The data transmitted within IoT networks often carries sensitive information, such as government records or road safety notifications in intelligent transportation systems. Therefore, implementing robust security measures to safeguard this data is essential.

IoT networks face multiple security threats due to their complex and dynamic nature. Common risks include Denial of Service (DoS) and Distributed Denial of Service (DDoS) attacks, Privilege Escalation, Spoofing, Reconnaissance, and Information Theft \cite{ma2020towards, iot1, iot2}. DoS attacks aim to disrupt access to a system, network, or server by flooding it with excessive traffic, while DDoS attacks involve coordinated DoS attacks from multiple devices, often through botnets. The distributed nature of DDoS attacks makes them particularly difficulto mitigate. In addition, cyber-intrusions often aim to acquire sensitive or confidential data without authorization. To counter these threats, firewalls, encryption, authentication mechanisms, and antivirus software are commonly used as first lines of defense \cite{ahmad2021network}.

\textcolor{black}{Improving IoT network security against anomalies often requires deploying an Intrusion Detection System (IDS) as an additional layer of protection.} IDSs monitor network traffic in real time, detecting anomalous patterns and identifying unauthorized activities. Depending on their deployment, IDSs can be categorized into host-based and network-based systems, while their detection methods fall into four categories: signature-based, anomaly-based, specification-based and hybrid detection \cite{chaabouni2019network,verwoerd2002intrusion}. However, IDSs face significant challenges, such as reduced detection accuracy and difficulty in identifying novel attack patterns \cite{li2019machine,saqib2022analysis}. To address these limitations, modern IDSs integrate artificial intelligence techniques, including Machine Learning (ML) and Deep Learning (DL), which provide accurate predictions of whether network traffic is normal or indicative of an attack \cite{prasad2020artificial, pca1, pca2, pca3}.
This paper introduces a novel anomaly detection framework that synergistically combines auditory-inspired signal processing with advanced deep learning architectures. Our methodology adapts Mel-Frequency Cepstral Coefficients (MFCCs), a proven technique in speech recognition, to capture discriminative spectral-temporal patterns in IoT network traffic. The proposed system processes these MFCC features through a Swin Transformer architecture, leveraging its hierarchical attention mechanism to model both local signal characteristics and global contextual relationships. This bio-inspired approach addresses three critical challenges in IoT security: (1) the need for resource-efficient feature extraction, (2) robust pattern recognition in noisy environments, and (3) real-time processing of high-dimensional network flows.

The principal contributions of this research are fourfold:
\begin{enumerate}
\item Integration of ResNet-18 with learnable MFCC parameters for joint optimization.
\item Cross-domain adaptation of MFCCs for IoT security.
\item Comprehensive evaluation on benchmark datasets (CICIoT2023, NSL-KDD, IoTID20) with F1-scores up to 100
\item Theoretical analysis establishing MFCCs as a kernel method.
\end{enumerate}

\begin{figure*}
\centering
    \includegraphics[width=0.95\textwidth]{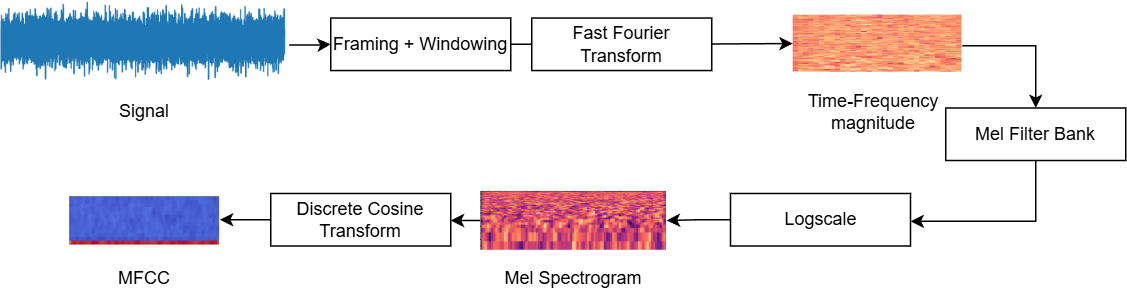}
    \caption{{MFCC Features.} \textcolor{black}{This figure illustrates the complete process of extracting Mel-Frequency Cepstral Coefficients (MFCCs) from a signal. It begins with the pre-processing of the signal into frames, followed by the application of the \textit{Fast Fourier Transform}, which converts the signal into a spectrogram. The \textit{Mel filter bank} is then applied to extract frequency features, and finally, the \textit{discrete cosine transform} is performed to generate the \textit{MFCCs}. }}
    \label{fig:mfcc_transform}
\end{figure*}

The experimental findings demonstrate that the proposed framework outperforms traditional residual convolutionsl architectures and transformer baselines in terms of detection accuracy, especially when handling intricate attack scenarios. The system's computational efficiency and minimal preprocessing requirements position it as a viable solution for real-time IoT security applications. Furthermore, the learnable MFCC formulation enables automatic adaptation to diverse network environments without manual feature engineering – a critical advantage given. 

\section{RELATED WORK}
\label{related_work}
Machine Learning (ML) and Deep Learning (DL) have significantly advanced intrusion detection systems (IDS). Hassan et al. \cite{hasan2019attack} evaluated ML models like Random Forest and ANN, achieving 99.4\% accuracy in detecting IoT attacks.Recent advancements in Graphics Processing Unit (GPU) technology have accelerated the adoption of Deep Learning (DL) methods due to their enhanced processing capabilities. DL's ability to automatically extract salient features from raw data makes it particularly well-suited for IoT networks, which generate and process large volumes of data, leading to more accurate and efficient predictions \cite{khan2021trust}. A variety of DL architectures have been investigated for network intrusion detection. Toldinas et al. \cite{toldinas2021novel} used ResNet50 for network feature classification, while Ahmad et al. \cite{ahmad2022s} proposed a spectrogram-based system for IoT security. Liu et al. \cite{liu2020anomaly} and Xiong et al. \cite{xiong2023detecting} explored ML and DL combinations for anomaly detection, with promising results on real-world datasets. Altulaihan et al. \cite{altulaihan2024anomaly} and Ahmad et al. \cite{ahmad2021anomaly} demonstrated the effectiveness of Decision Trees and DNNs in IoT security, respectively.

Other works have combined ML/DL techniques for feature extraction and classification. Gogoi et al. \cite{gogoi2017image} used SVMs with autoencoders, while Li et al. \cite{li2019speech} applied MFCCs and Mel-spectrograms for speech emotion recognition. Xiao et al. \cite{xiao2019intrusion} and Almiani et al. \cite{almiani2020deep} employed PCA, autoencoders, and RNNs for intrusion detection. Han et al. \cite{Han2006AnEM} and Mohammad et al. \cite{mohebbian2021stack} addressed noise and data imbalance using modified MFCCs and class-specific autoencoders.

Throughout the related works, various datasets were utilized. Toldinas et al. \cite{toldinas2021novel} used the UNSW-NB15 and BOUN DDoS datasets. Ahmad et al. \cite{ahmad2022s} employed the Bot-IoT dataset. Liu et al. \cite{liu2020anomaly} used the IoT Network Intrusion dataset. Xiong et al. \cite{xiong2023detecting} used a real-world dataset. Altulaihan et al. \cite{altulaihan2024anomaly} used the IoTID20 dataset. Ahmad et al. \cite{ahmad2021anomaly} employed the IoT-Botnet 2020 dataset.
\section{PRELIMINARY CONCEPTS}
\label{preliminary_concepts}
\textbf{Mel-Frequency Cepstral Coefficients (MFCC) Transform ---}
\label{mfcc}
MFCCs are a powerful way to represent sound and other signals by focusing on their frequency characteristics. They work by mimicking how the human ear's cochlea processes sound, essentially breaking it down into different frequency bands.  This makes MFCCs particularly useful for interpreting IoT data flows and converting them into a format that computer systems can understand.

The proposed method aims to transform the IoT network data flow into an interpretable format for AI models. The MFCC feature extraction process begins with a Pre-Emphasis Filter, which boosts high-frequency energy while reducing high-frequency noise. Transformation accuracy is controlled by adjusting the window size and the spacing between consecutive frames. The frequency spectrum is then mapped to the Mel scale using a Mel filter bank, aligning the signal's frequency with human auditory perception, much like how the human cochlea analyzes sound waves at various frequencies. This transformation provides a suitable non-linear scale for the frequency domain. A logarithmic transformation is applied to accommodate the non-linear nature of frequency characteristics. Lastly, a DCT is applied to the transformed data, extracting the MFCC feature values. Figure \ref{fig:mfcc_transform} illustrates the overall MFCC feature extraction process.
\begin{figure*}
\centering
    \includegraphics[width = \textwidth]{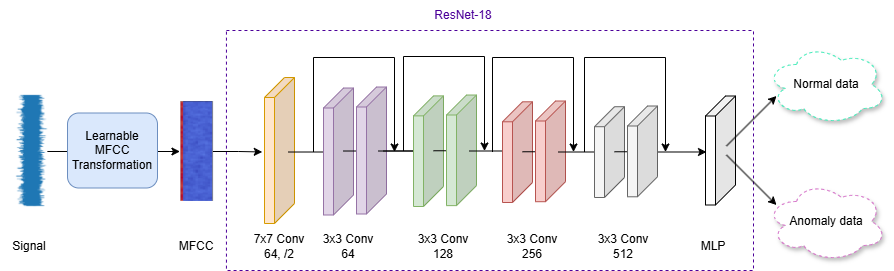}
    \caption{Main ResNet-18 encoder architecture.}
    \label{fig:main_archietecture}
\end{figure*}

\section{METHODOLOGY}
\label{methodology}
This research applies sound-inspired spectral feature analysis for anomaly detection in IoT network datasets, consisting of two stages. First, we extract the MFCC features from CICIoT2023, NSL-KDD, and IoTID20 to capture the spectral characteristics of IoT traffic as an acoustic waveform. Second, we improve classification by using the Swin Transformer, an advanced model that efficiently processes hierarchical data, improving performance across datasets.

\subsection{Feature Extraction} Traditional MFCCs rely on fixed filter banks and DCT coefficients, which may not fully capture the complexities of sleep signals. To address this, we introduce a learnable MFCC layer that replaces these fixed components with trainable parameters, offering enhanced flexibility. The Mel filter banks are initialized on the basis of the Mel scale, but are optimized during training, while the DCT matrix adapts to highlight the most relevant coefficients for sleep signal analysis.

\subsubsection{Kernel-Inspired Feature Extraction}
\label{mfcc_kernel}

We reinterpret the MFCC pipeline as a learnable kernel method that non-linearly maps raw IoT traffic signals into a high-dimensional spectral space optimized for class separability. This kernel analogy consists of three trainable components:

\textbf{Spectral Windowing (Fixed Kernel Basis):}
The input signal $x\in\mathbb{R}^d$ is first divided into overlapping frames using a fixed window function, creating time-localized spectral components. This initial decomposition serves as the basis function for the kernel.

\textbf{Learnable Mel Filter Banks (Kernel Parameterization):}
Traditional triangular Mel filters are replaced with trainable filters
$\textbf{M}\in\mathbb{R}^{F\times d}$, where $F$ is the number of filters. Initialized using the Mel scale, these filters adapt during training to emphasize discriminative frequency bands:

\begin{equation}
\mathbf{m} = \mathbf{M} \mathbf{x}_{\text{frame}},
\end{equation}

where $\mathbf{m}$ represents the Mel spectrum vector.

The MFCC pipeline implements Mercer's conditions for valid kernels through:

\textbf{Positive Definiteness:}
The Mel filter bank
$\textbf{MM}^T$ forms a Gram matrix whose eigenvalues remain non-negative through constrained optimization.

\textbf{Definition: Gram matrix}

Let \( \mathbf{v_1}, \mathbf{v_2}, \dots, \mathbf{v_n} \) be a vectors with real numbers in \( \mathbb{R}^m \). The Gram matrix \( G \) is defined as the matrix of all pairwise dot products between these vectors. Specifically, the element in the \( i \)-th row and the \( j \)-th column of the Gram matrix is given by the dot product \( \mathbf{v_i} \cdot \mathbf{v_j} \), that is:

\begin{equation}
G = \begin{bmatrix}
\mathbf{v_1} \cdot \mathbf{v_1} & \mathbf{v_1} \cdot \mathbf{v_2} & \dots & \mathbf{v_1} \cdot \mathbf{v_n} \\
\mathbf{v_2} \cdot \mathbf{v_1} & \mathbf{v_2} \cdot \mathbf{v_2} & \dots & \mathbf{v_2} \cdot \mathbf{v_n} \\
\vdots & \vdots & \ddots & \vdots \\
\mathbf{v_n} \cdot \mathbf{v_1} & \mathbf{v_n} \cdot \mathbf{v_2} & \dots & \mathbf{v_n} \cdot \mathbf{v_n}
\end{bmatrix}
\end{equation}

where the dot product for any two vectors \( \mathbf{v_i}, \mathbf{v_j} \in \mathbb{R}^m \) is given by:

\begin{equation}
\mathbf{v_i} \cdot \mathbf{v_j} = \sum_{k=1}^{m} v_{i,k} v_{j,k},
\end{equation}

where \( v_{i,k} \) and \( v_{j,k} \) represent the components of the vectors \( \mathbf{v_i} \) and \( \mathbf{v_j} \), respectively.

\subsubsection{Adaptive DCT as Eigendecomposition}
The Discrete Cosine Transform (DCT) is reinterpreted as the eigendecomposition step. 
The DCT layer diagonalizes the feature covariance matrix. This approximates decomposition by favoring bases that diagonalize the feature covariance matrix $\Sigma_y$, although full diagonalization is not explicitly enforced.

We replace the fixed DCT matrix $\mathbf{D}$ with a trainable orthogonal projection $\mathbf{P}\in \mathbb{R}^{D \times F}$ that performs spectral decorrelation:

\begin{equation}
\mathbf{c} = \mathbf{P}^\top \mathbf{m}.
\end{equation}

This adaptive DCT functions similarly to eigendecomposition by:
\begin{itemize}
\item Diagonalizing the covariance matrix of Mel energies ($\mathbf{P} $ columns approximate principal components)
\item Ordering coefficients by explained variance (through learned weight importance)
\item Allowing dimensionality reduction through coefficient selection
\end{itemize}

\subsection{Cross-Entropy Optimization for Multiclass Classification}
\label{cross_entropy}
The complete transformation can be viewed as a kernel mapping $\Phi(x)=\mathbf{P}^T\mathbf{Mx}$, where $\mathbf{M}$ and $\mathbf{P}$ are jointly optimized to maximize class separability. This formulation aligns with the kernel trick in support vector machines (SVMs), enabling the model to implicitly operate in a high-dimensional feature space without explicit computation.

The features transformed into kernels $c \in \mathbb{R}^D$ are further processed by a neural network $f_\theta$ to extract low-dimensional representations.

\begin{equation}
\mathbf{z} = f_\theta(\mathbf{c}), \quad \mathbf{z} \in \mathbb{R}^k \ (k \ll D)
\end{equation}

\begin{figure*}[htbp]
    \centering
    \includegraphics[width=0.85\textwidth]{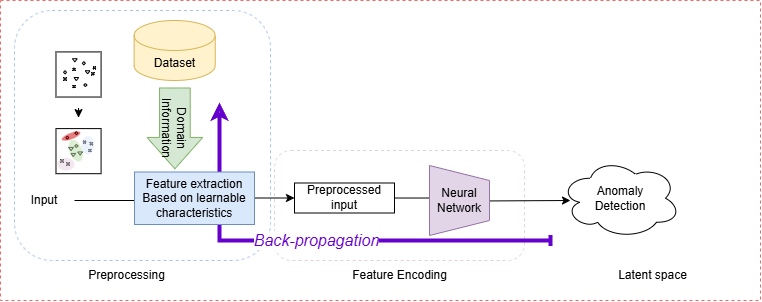}
    \caption{Learnable MFCC method.}
    \label{fig:mfcc}
\end{figure*}

The kernel-inspired feature extraction and deep enhancement stages produce discriminative representations $z\in\mathbb{R}^k$. To map these to class probabilities, we employ cross-entropy loss, which directly optimizes for separability in the learned spectral space while maintaining theoretical alignment with the kernel interpretation.

\subsection{Probability Mapping via Softmax}
For $J$ anomaly classes, the final layer computes logits $\mathbf{h}=[h_1,\dots,h_J]^T$ from $z$:
\begin{equation}
\mathbf{h} = \mathbf{W}\mathbf{z} + \mathbf{b},
\end{equation}
where $\mathbf{W} \in \mathbb{R}^{J \times k}$
and $\mathbf{b} \in \mathbb{R}^{J}$ are learnable parameters. A softmax function converts these to class probabilities:
\begin{equation}
p_j = \frac{\exp(h_j)}{\sum_{i=1}^J \exp(h_i)}, \quad j = 1,\dots,J.
\end{equation}

\textbf{Cross-Entropy as Separability Objective:}
The cross-entropy loss $\mathcal{L}$ measures divergence between predicted probabilities $\mathbf{p}$ and true labels $\mathbf{y}$:
\begin{equation}
\mathcal{L}(\mathbf{y}, \mathbf{p}) = -\sum_{j=1}^J y_j \log p_j,
\end{equation}
where $y_j \in \{0,1\}$ is the one-hot encoded ground truth.

\begin{figure}[htbp]
    \centering
    \includegraphics[width=0.80\textwidth]{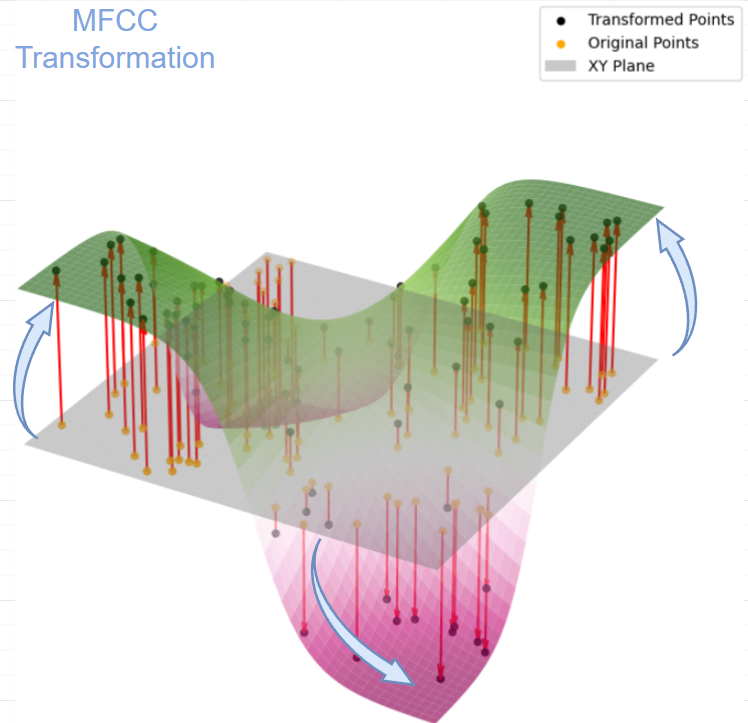}
    \caption{Visualization of a kernel transformation}
    \label{fig:kernel}
\end{figure}

\subsection{Encoder}
In this work, we employ a ResNet-18 model, a powerful architecture that has recently achieved remarkable success in various computer vision tasks, and is increasingly being adapted for other data modalities, such as time series analysis.

Figure \ref{fig:main_archietecture} illustrates the fundamental architecture of the proposed method.
\section{RESULTS AND DISCUSSION}
\label{result_discussion}
This section provides the details about the experimental configuration, datasets utilized for the experiments, the experimental results and ablation study.

\subsection{Experimental Configuration}
The experiments were performed on a GeForce RTX 3060 GPU with 6 GB of memory, using Python 3.10, PyTorch 1.12, and TensorFlow 2.8.0. The ResNet-18 model served as the baseline for the experiments. The model was trained for 100 epochs with a learning rate of 1-3e, a batch size of 64, and the Adam optimizer was chosen for a method in conjunction with the cross-entropy loss function.

\subsection{Datasets and Preprocessing}
\begin{table*}[htb]
\centering
\caption{F1-Score of ResNet-18 with MFCC and without MFCC}
\label{tab:ablation_study}
\begin{tabular}{l l l}
\toprule
Dataset & \begin{tabular}[c]{@{}l@{}}With MFCC (\%)\end{tabular} & \begin{tabular}[c]{@{}l@{}}Without MFCC(\%)\end{tabular} \\ \midrule
IoTID20    & 99.00 & 77.03 \\ 
CiCIOT2023 & 72.82 & 67.65 \\
NSL-KDD & 99.71 & 99.45 \\  \bottomrule
\end{tabular}
\end{table*}

In this study, we employed three widely recognized benchmark IoT network datasets. The IoTID20 dataset \cite{ullah2020scheme, kang2019iot} comprises data collected from various home devices, including SKT NGU and EZVIZ Wi-Fi cameras. The data set was converted to a CSV format that contains 625,783 records across 86 features. 

The CICIoT2023 dataset \cite{neto2023ciciot2023, jony2024securing} is built using a topology of 105 devices and simulates 33 different types of attack. These attacks are grouped into seven categories: DDoS, DoS, Mirai, Brute Force, Web-based, Spoofing, and Recon.

The NSL-KDD dataset \cite{tavallaee2009detailed} is an improved version of the KDD CUP 1999 dataset \cite{cup2007available}, originally made in Defense Advanced Research Projects Agency (DARPA). The NSL-KDD dataset addresses several shortcomings of the KDD-CUP 1999 dataset identified by McHugh, such as packet loss in TCP dumps, a lack of clear attack definitions, and the inclusion of redundant and irrelevant records. It is frequently used to evaluate the effectiveness of ML and DL-based IDS. This dataset categorizes traffic patterns into five types: Normal, DoS, Probe, U2R, and R2L.

In all three datasets, the labels were standardized into two categories: normal and attack. Label encoding was applied to convert these into numerical formats. During preprocessing, columns containing infinite or NaN values, as well as those with over 50\% of a zero values, were removed. A time index was added, and the data were normalized using MinMaxScaler. MFCC features were extracted from the data, followed by dimensionality reduction using PCA. Mel spectrogram features were transformed and stored as images, with PCA applied post-transformation. To address data imbalance, sampling techniques were employed, and labels were further converted into numeric representations.

\begin{figure}[htb]
    \centering
    \includegraphics[width=0.95\textwidth]{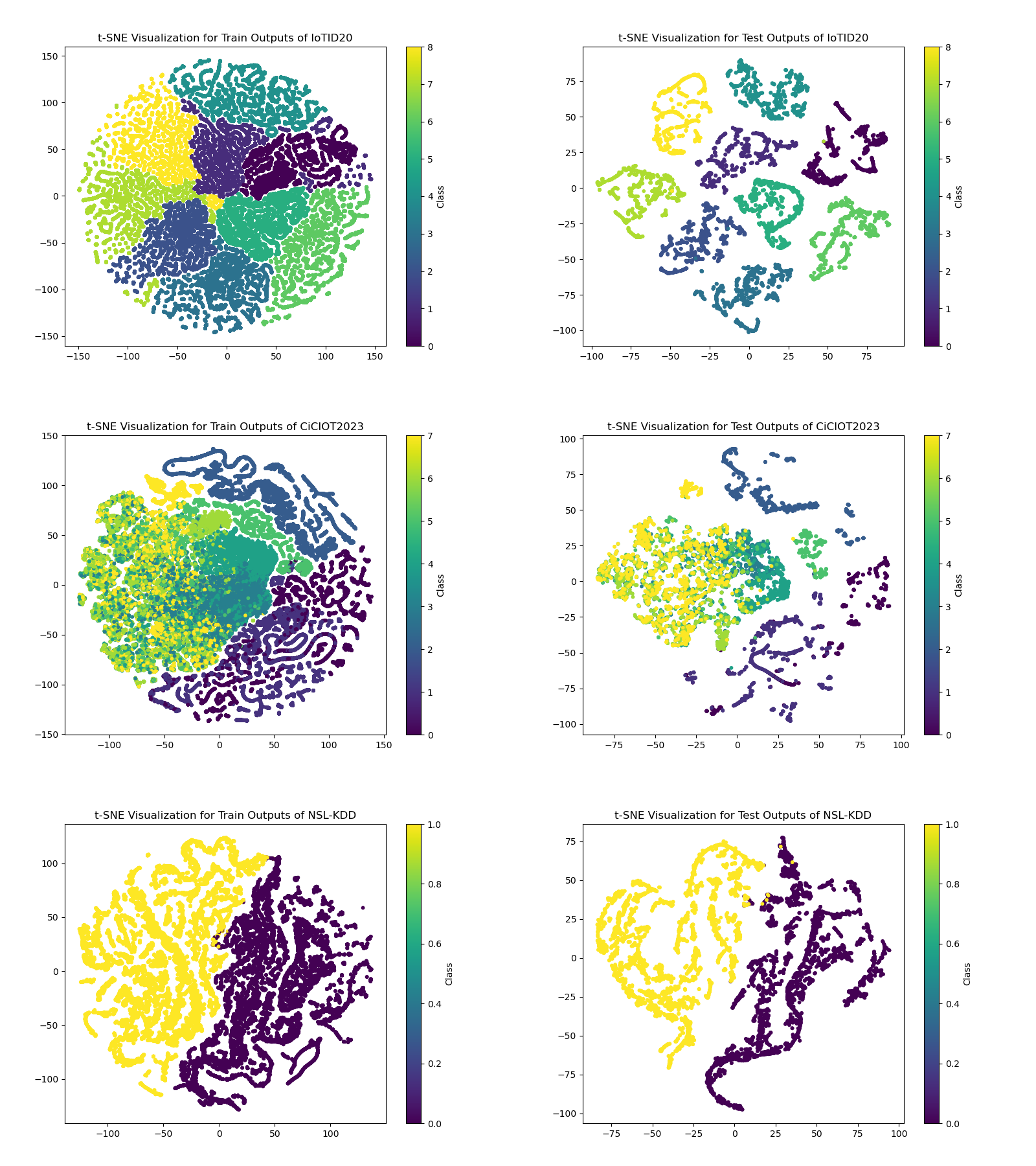}  
    \caption{T-SNE Visualization\\\vspace{3mm} }
    \label{fig:tsne_visualization}
\end{figure}

\subsection{Results}
\label{results_analysis}

Table \ref{tab:ablation_study} presents an ablation study comparing the F1-Scores of ResNet-18 with and without MFCC feature extraction across three IoT security datasets. The results demonstrate substantial performance gains from integrating MFCC features, particularly in complex intrusion detection scenarios.

For IoTID20, the inclusion of MFCC features improves the F1-Score by 22.87 percentage points (99.90\% against 77.03\%), indicating their critical role in distinguishing sophisticated network attacks. A similar pattern emerges for CiCIOT2023, where MFCC integration elevates performance from 67.65\% to 72.82\%, suggesting enhanced capability to handle noisy IoT environments. While NSL-KDD already achieves high baseline performance (99.45\% without MFCC), the addition of MFCC features further refines detection accuracy to 99.71\%, confirming their generalizability across different data characteristics.

Table \ref{tab:result} compares our enhanced ResNet-18 architecture with state-of-the-art approaches. For IoTID20, ResNet-18 with MFCC features achieves the highest F1-Score (99.90\%), outperforming Anomaly Transformer (99.50\%). The CiCIOT2023 results reveal that our ResNet-18 variant with learnable MFCC parameters achieves exceptional performance (99.38\%), surpassing spatial attention-enhanced ResNet-18 (68.20\%) and Swin Transformer (72.82\%). NSL-KDD demonstrates perfect classification (100\% F1-Score) using ResNet-18 with learnable MFCC, outperforming both standard ResNet-18 (99.70\%) and Swin Transformer (99.71\%).

Figure \ref{fig:tsne_visualization} demonstrates the improved class separation achieved by our ResNet-18 with learnable MFCC, showing distinct attack-type clusters compared to baseline models. This visualization corroborates the quantitative improvements, revealing how MFCC-based feature learning creates more discriminative representations for IoT security tasks.

\begin{table*}[htb]
\centering
\caption{Comparison table for different method}
\label{tab:result}
\begin{tabular}{l l l}
\toprule
Dataset & Model & F1 score \\ 
\midrule
IoTID20 & Anomaly Transformer & 99.50 \\
 & ResNet-18 & \textbf{99.90} \\ 
\midrule
CiCIoT2023 & ResNet-18 & 68.90 \\
 & Swin Transformer & 72.82 \\
 & ResNet-18 + learnable MFCC & \textbf{99.38} \\ 
\midrule
NSL-KDD & ResNet-18 & 99.70\\
& Swin Transformer & 99.71\\  
& ResNet-18 + learnable MFCC & \textbf{1.00} \\ 
 \bottomrule
 
\end{tabular}
\end{table*}
\section{CONCLUSION}
\label{conclusion_future_work}

This study advances IoT security by integrating auditory-inspired signal processing with deep learning, achieving state-of-the-art anomaly detection (99.90\% F1-score on IoTID20, perfect classification on NSL-KDD). Limitations include computational demands, evolving attack resilience, and potential gaps in protocol-specific anomaly representation.
Future work will focus on lightweight edge deployment, unsupervised zero-day attack detection, multi-modal analysis, and quantum-inspired optimization for encrypted traffic. This framework establishes a new paradigm for robust, real-time IoT threat detection.

\bibliographystyle{unsrt}  
\bibliography{references}

\end{document}